\documentclass[11pt,a4paper,english]{article}

\usepackage{graphicx}
\usepackage{dcolumn}
\usepackage{bm}
\usepackage{natbib}

\begin{document}



\title{A Laplace Transform Method for Molecular Mass Distribution Calculation from Rheometric Data} 

\author{C. Lang}





\date{\today}



\maketitle 
\section*{Abstract}
Polydisperse linear polymer melts can be microscopically described by the tube model and fractal reptation dynamics, while on the macroscopic side the generalized Maxwell model is capable of correctly displaying most of the rheological behavior. In this paper, a Laplace transform method is derived and different macroscopic starting points for molecular mass distribution calculation are compared to a classical light scattering evaluation. The underlying assumptions comprise the modern understanding on polymer dynamics in entangled systems but can be stated in a mathematically generalized way. The resulting method is very easy to use due to its mathematical structure and it is capable of calculating multimodal molecular mass distributions of linear polymer melts.
\section{\label{1} Introduction}

The linearized macroscopic behavior of polymer solutions and melts under shear deformation can be described by the generalized N-mode Maxwell model \cite{EvMo2008}. It is known that the Fourier transform of stress and strain rate in the high frequency regime as well as that of stress and shear rate in the low frequency regime are connected via two different material properties (MPs), namely, the shear modulus and the shear viscosity. Those two quantities are interdependent and both depend on the molecular mass distribution.

In the quasi-linear regime, the generalized Maxwell model in Fourier space possesses two measurables, the storage- and the loss modulus. Those two quantities can be used to calculate the non-linear complex viscosity. By applying the Cox-Merz relation, the linearized behavior can, furthermore, be used to determine the non-linear shear viscosity as a func tion of the shear rate, which can be thought of as a third measurable.

From the microstructural side,  polymer dynamics are used to find an appropriate description for the stress in the liquid in order to be able to connect the micro- and macroscopic view \cite{DoEd1986}. Here, the classical N-particle Smoluchowski equation approach is chosen, describing the probability state of the many chain system by an equation of motion for the probability density function of particle position. However, when only monodisperse linear polymer melts are considered, their probabilistic treatment can be simplified drastically. The Smoluchowski equation for the single reptation model inside the tube simplifies to a one dimensional diffusion equation which is solved to calculate appropriate autocorrelation functions leading to the molecular weight dependence of the stress tensor. Taking into account the polydisperse nature, however, is only feasible by superimposing a distribution of probablistic reptation modes onto the original description or by solving a more elaborated form of the Smoluchowski equation. Here, as is discussed in section \ref{5.1}, the first approach is chosen, since it has been shown to be very versatile and give good results in earlier, related works \cite{Ande1998}, \cite{Thim1999}.

With the combination of those two models, it can be shown that the viscoelastic properties, or as they are called here: the MPs, of polydisperse melts can be used to directly calculate the molecular mass distribution (MMD) or vice versa. Manifold approaches like this are readily found in the literature \cite{CaGu1997}, \cite{NoCo2001}, \cite{NoCo+1996}, \cite{MaTe1987}, \cite{MaTe1988}, \cite{RuKe+2005} and most recently in a paper by Pattamaprom et al. \cite{PaLa+2008}. Nonetheless, no general treatment has been given yet. Here, it will be shown that it is mathematically near to irrelevant which MP is used to find the MMD. The shown procedure is very simple and can be compared to the existing approach of MMD calculation from light scattering detection \cite{Gresc1981}.

\section{\label{2} Theory}

As a starting point, one can take an analytical result by Thimm et al. \cite{Thim1999}, specifying the relation between the molecular mass distribution $\psi(m)$ and the continuous relaxation time spectrum $h(\tau)=h(\tau(m))=\tilde{h}(m)$ :
\begin{equation}
\label{ThimmTheo}
\psi(m)=\lambda \tilde{h}(m)~~,
\end{equation}
 where one uses the connection between molecular mass $m$ and longest relaxation time $\tau$ of a polymer species:
\begin{equation}
\label{TauM}
\tau=\zeta m^\alpha~~,
\end{equation}
which is mathematically the simplest direct interdependence of those two quantities. The validity range of this relation, however, can be disputed in view of the broad range of existing polymeric materials. Especially for the case of immiscible blends 
 and highly branched polymers \cite{DePe+1993}, equation~\ref{TauM} can loose its validity, while it is supposed to hold for most uni- and bimodal polydisperse linear polymer melts \cite{ReVi+2000}.
One, thereby, defines:
\begin{equation}
\label{lambda}
\lambda=\frac{1}{\beta}\left(\frac{\alpha}{G_N^0}\right)^{\frac{1}{\beta}}\left[\int_{m_e}^\infty dm m^{-1}\tilde{h}(m)\right]^{(1-\beta)/\beta}~~,
\end{equation}
where $\beta$ can be called the fractal reptation exponent, $G_N^0$ is the plateau modulus and $m_e$ the entanglement molecular mass.

In order to develop this result further, one can use an approximation of the molecular mass distribution. A very useful function for this purpose is the generalized exponential (GE) function, which has found use in many earlier papers on the matter \cite{Gros1988}, \cite{CaGu1997}, \cite{NoCo2001}, \cite{CoNo2003}, \cite{NoCo+1996}. However, it has been overlooked that the use of a single GE distribution is a strong limitation of the accuracy of approximation, especially for the case of long molecular weight tails. Instead, one can choose the N-mode GE distribution:
\begin{equation}
\label{GEX}
\psi(m)=\frac{1}{N}\sum_{i=1}^N\gamma_im^{a_i}\exp[-b_im^{c_i}]~~,
\end{equation}
where $\gamma$ is the normalization constant for a single mode and only $a$ and $c$ are independent parameters. This is equivalent of saying that the molecular mass possesses a distribution at all. As a matter of fact, this function is applicable for the approximation of a much larger span of functions than one will find as naturally occuring MMDs \cite{VaHe1984}. Using this approximation together with some basic definitions within the generalized Maxwell model leads to a full description of linear and nonlinear rheological quantities in terms of the MMD.

At first, one can find the shear modulus $G(t)$ by using the inverse form of equation~\ref{ThimmTheo} together with equation~\ref{GEX}:
\begin{equation}
\label{ShearModulus}
G(t)=\int_0^\infty d\tau\tau^{-1} h(\tau)exp\left[-\frac{t}{\tau}\right]=\frac{1}{N}\sum_{i=1}^N\hat{\lambda}_i\Gamma(\xi_i)\mathcal{L}_{\xi_i}(t,0)[g_i(\tau)]~~,
\end{equation}
where 
$$\mathcal{L}_\lambda(s,\alpha)[f(x)]=\frac{1}{\Gamma(\lambda)}\int_\alpha^\infty dx x^{\lambda-1}f(x)\exp[-sx]$$
is called the incomplete Laplace transform \cite{Temm1987}, $\Gamma(\lambda)$ is the gamma function, $\hat{\lambda}=\frac{\gamma\zeta^{\xi}}{\lambda}$, $\xi=-\frac{a}{\alpha}$, $g(\tau)=\exp\left[-k\tau^\mu\right]$, $k=b\zeta^{-\mu}$ and $\mu=\frac{c}{\alpha}$.

This leads directly to the measurables in Fourier space, the storage modulus $G'(\omega)$ and the loss modulus $G''(\omega)$:
\begin{equation}
\label{gprime}
G'(\omega)=\frac{1}{N}\sum_{i=1}^N\frac{\hat{\lambda}_i}{\mu_i}\Gamma(\hat{\xi_i})\mathcal{L}_{\hat{\xi_i}}(k_i,0)[\hat{f}_i'(\tau,\omega)]~~,
\end{equation}
\begin{equation}
\label{gdprime}
G''(\omega)=\frac{1}{N}\sum_{i=1}^N\frac{\hat{\lambda}_i}{\mu_i}\Gamma(\hat{\xi_i})\mathcal{L}_{\hat{\xi_i}}(k_i,0)[\hat{f}_i''(\tau,\omega)]~~,
\end{equation}
where $\hat{\xi}=\frac{1}{\mu}$,
\begin{equation}
\hat{f}'(\tau,\omega)=\frac{\omega^2\tau}{1+(\tau\omega)^2}
\end{equation}
and 
\begin{equation}
\hat{f}''(\tau,\omega)=\frac{\omega\tau^{2}}{1+(\tau\omega)^2}~~.
\end{equation}
Using these two relations, one can write down the quasi-nonlinear form of the shear viscosity: 
\begin{equation}
\label{viscosity}
|\eta^\ast(\omega)|=\frac{1}{\omega}\sqrt{\left(G'(\omega)\right)^2+\left(G''(\omega)\right)^2}~~.
\end{equation}
 By applying the Cox-Merz rule this function is comparable to the shear rate dependent nonlinear viscosity.

\section{\label{3}Materials and Measurements}
\subsection{\label{3.1} Materials}

\begin{figure}[h]
	\begin{center}
		\includegraphics[scale=0.5]{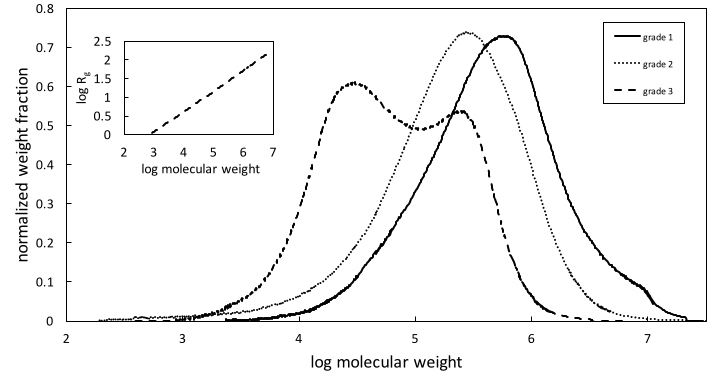}
		\caption{$d\psi/dlog(m)$ vs. $log(m)$ for the three polymer grades. Inlet: Radius of gyration as a function of molecular weight for grade 3.}
		\label{fig6}
	\end{center}
\end{figure}

For providing a broad range of MMDs, 2 linear polymer types of Borealis AG were chosen which differ in composition and MMD. Additionally, one polymer from Dow was chosen as an example for a bimodal grade. The polymers are listed in table~\ref{tab1}. Product data sheets of the grades are readily available from Borealis AG \cite{Bore0000} as well as Dow \cite{Dow0000}. 
The molecular mass moments $\overline{M_w}= \sum_i{n_im_i}/\sum_i{n_i}$ and $\overline{M_n}= \sum_i{w_im_i}$, where $n_i$ is the number of molecules with molecular mass $m_i$ and $w_i=n_im_i/\sum{n_im_i}$, from triple detection gel permeation chromatography are given in table~\ref{tab4}. In figure~\ref{fig6}, the measured MMDs are shown for all 3 grades. It is seen that grade 2 and 1 can be mainly distinguished by their high molecular mass tail, which is well pronounced for grade 2.

Grade 3, in addition to its bimodality, has a low degree of chain branching with octene sidegroups, as is indicated by the viscosity contraction factor $g'=(R_g^{br}/R_g^{lin})^\epsilon$, where $R_g^{br}$ is the radius of gyration of the branched polymer and $R_g^{lin}$ that of the linear polymer and $\epsilon$ is the structure factor, taken as $\epsilon=0.75$. Since the viscosity contraction factor as well as the structure factor scale with the molecular weight \cite{BeCa+2000}, $g'$ is evaluated at a molecular weight of $10^6$~g/mol, where the change of $g'$ with $m$ is expected to be relatively moderate. The viscosity contraction factor for the high molecular weight tail of the bimodal grade is $g'=0.93$ which corresponds to a number of 10 branches per $10^4$ C atoms, therefore, the degree of chain branching is low. This is furthermore indicated by the logarithmic linearity of the radius of gyration as a function of molecular mass, see the caption in figure~\ref{fig7}.

\begin{table}[h]
	\caption{Materials}
	\smallskip{}
	\centering{}
	\begin{tabular}{||c|c|c|c||}
		\hline
		\textbf{abbreviation} & \textbf{name}& \textbf{producer}& \textbf{typical application}\\
		\hline
		grade 1&PP RA130E& Borealis& extrusion, pipes \\
		grade 2& PP HC205TF&Borealis&extrusion, thermoforming \\
		grade 3 & PE RT2388 &Dow & extrusion, pipes \\
		\hline
	\end{tabular}
	\label{tab1}
\end{table}

\begin{table}[h]
	\caption{Materials}
	\smallskip{}
	\centering{}
	\begin{tabular}{||c|c|c||}
		\hline
		\textbf{grade} & $\overline{M_w}$& $\overline{M_n}$\\
		 & g/mol & g/mol \\
		\hline
		 1&207000& 393000 \\
		 2& 88577&279147 \\
		 3 & 31158 &65189 \\
		\hline
	\end{tabular}
	\label{tab4}
\end{table}

\subsection{\label{3.2} Measurements}

In order to compare the results from the theoretical description given in section~\ref{2} with the widely used light scattering measurement for MMD determination, a high-temperature gel permeation chromatography (HT-GPC) with subsequent triple detection has been carried out in a Viscotec HT-GPC system at the institute of polymer chemistry at the Johannes Kepler University in Linz, Austria. The measurement temperature was 140~$^o$C in a threefold column. 20 mg of each polymer in table~\ref{tab1} were dissolved in 10 ml 1,2,4-trichloro-benzene and stored at 140~$^o$C for 45 min. The measurement was calibrated with a polystyrene-standard of 99~kDa molecular weight. Data where analyzed using the OmniSEC GPC software.

The linearized rheological data were measured at the institute of polymer extrusion and building physics at the Johannes Kepler University Linz. A strain controlled measurement was conducted at an Anton Paar MCR 302 rheometer with a cone-plate geometry of 1.5~$^o$ cone angle and 25~mm diameter at measurement temperatures of 190~$^o$C and 200~$^o$C. For this purpose, 25~mm diameter tabs were pressed from granular stock by using a H\"ofer hydraulic laboratory press at 200~$^o$C. Data were analyzed in the Anton Paar rheometer software.
Nonlinear viscosity data were measured in a slit capillary die from Thermo Scientific at the institute of polymer extrusion and building physics at the Johannes Kepler University Linz, which was connected to a gear pump after the Haake Rheomex 19/33 OS single screw extruder from Thermo Scientific \cite{Ther2016}. The slit geometry was 125 x 20 x 1 mm. The measurement temperatures were set to 190~$^o$C and 200~$^o$C with a gear pump rotational speed from 1 to 400 $s^{-1}$. The chamber volume of the pump was 2.4 cm$^3$. Using the Weissenberg-Rabinowitsch correction after applying a quasi-Newtonian die curve approximation to the pressure-throughput data results in the desired viscosity curves. The pressure was, thereby, measured at 4 positions along the capillary while the throughput is set by the chamber volume of the gear pump and the rotational speed of this force feeding device. The used extruder speed was 200 rpm with a standard 3 zone screw of 19.05 mm diameter.

\section{\label{4} Application}

The equations from section~\ref{2} provide a consistent method for the calculation of MMDs from rheometric data. Since the MMD is approximated by the N-mode GE distribution with 2 independend parameters per mode, one has to solve at least a numerical fitting method with 2N free parameters. In general, the values for the fractal reptation exponent $\beta$, the molecular friction coefficient $\zeta$, the relaxation time exponent $\alpha$ and the Thimm \cite{Thim1999} coefficient $\lambda$ can also be taken as free parameters. This makes in total 2N+4 fitting parameters. Since this number quickly gets too large to be solvable in a meaningful way, it is recommended to measure $\alpha$, $\beta$ and $\zeta$ or take their values from the literature. It has to be said here that the literature values for $\zeta$ have to be taken very carefully.

\begin{figure}[h]
	\begin{center}
		\includegraphics[scale=0.5]{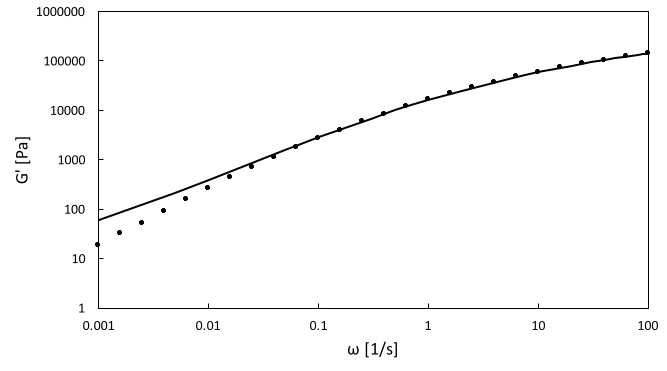}
		\caption{Storage modulus vs. frequency of grade 1 in comparison to backfitting.}
		\label{fig1}
	\end{center}
\end{figure}

In the numerical part of this study, the values for $\alpha$, $\beta$ and $\zeta$ where chosen according to table~\ref{tab2}. The Thimm coefficient was found by fitting the Cole-Cole model \cite{CoCo1942} to the measurement of $G'(\omega)$ for calculating an approximation for $h(\tau)$ \cite{Lang2016}. In order to check the validity of the approximation, the relaxation time spectrum from fitting was used to calculate back to $G'(\omega)$ and compare it to the measurement. The result is shown in figure~\ref{fig1}, where the line represents the back-fitted storage modulus calculated from the relaxation time spectrum which results from an inversion of the Cole-Cole fit to the measurement. It can be seen that the two curves agree well for higher frequencies, but the back-fitted result shows a discrepancy in the low frequency regime. This is attributable to the constraint-free extension of the Cole-Cole function in the low frequency regime, but due to the integral in equation~\ref{lambda}, this regime contributes only marginally to the coefficient $\lambda$.

\begin{table}[h]
	\caption{Material Constants}
	\smallskip{}
	\centering{}
	\begin{tabular}{||c|c|c||}
		\hline
		\textbf{name} & \textbf{value}& \textbf{source}\\
		\hline
		$\alpha$ &3.4& Schausberger \cite{Scha1991} \\
		$\beta$ & 3.84& see section~\ref{5.1} \\
		$\zeta$ & 3x$10^{-21}$ & Eder et al. \cite{Eder1989} \\
		\hline
	\end{tabular}
	\label{tab2}
\end{table}

\begin{figure}[h]
	\begin{center}
		\includegraphics[scale=0.5]{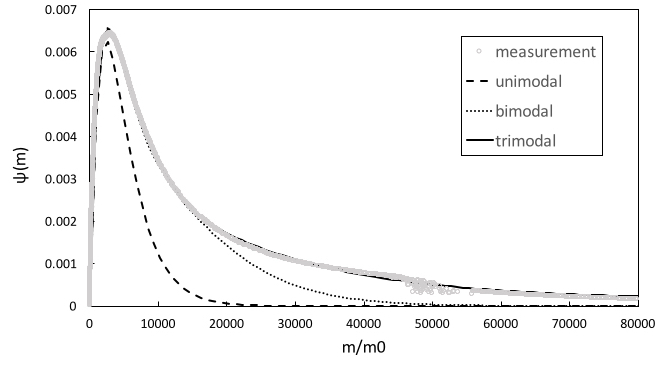}
		\caption{Uni- bi- and trimodal MMD calculation in comparison with HT-GPC measurement for grade 1 with $m_0=42.08$ Da.}
		\label{fig2}
	\end{center}
\end{figure}

Using those coefficients, the MMD of polydisperse linear polymers can be calculated by fitting the given functions for the measurables, equations~\ref{gprime} and \ref{gdprime}, to the measurements. It can be shown that the unimodal GE distribution does not approximate the HT-GPC measurement for grade 1 very well, while a subsequent increase in modes produces the full MMD accurately, including also the high molecular weight tail, figure~\ref{fig2}. In this case, it is already sufficient to include 3 modes for the GE distribution. The spacing of modes was chosen such that $\gamma_1<\gamma_2<\gamma_3$, which is of course inverse proportional to the contribution of the modes.
The calculation was done by using the Levenberg-Marquardt algorithm\cite{Leve1944}$^{,}$ \cite{Marq1963} and constraining the parameters $a_i>0$ and $c_i>0$, where $i=1,2,3$.

\begin{figure}[h]
	\begin{center}
		\includegraphics[scale=0.5]{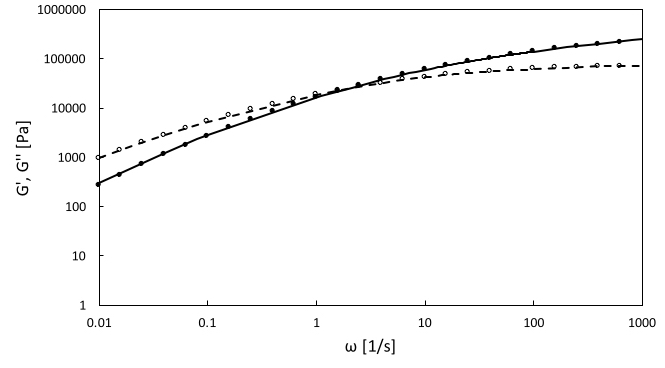}
		\caption{Model fit to measured G' (full dots), G'' (open circles)for grade 1 with.}
		\label{fig3}
	\end{center}
\end{figure}

For comparison, the linearized rheological measurements and their fits by the given model are shown in figure~\ref{fig3} for the case of grade 1. The quality of the fit strongly determines the resulting MMD. It should be emphasized here that the best method for using the given model is setting the known physical parameters and then step by step enhancing the number of modes of the GE function until the desired result is reached. In an automatically performed calculation, the number of modes can be set to 4 always, this is sufficient for describing most uni-and bimodal MMDs. In doing so, the calculation procedure is equivalent of inverting Laplace transforms and, thus, can be performed very easily. Here, the inverse Laplace transforms were calculated with Wolfram Mathematica.

\begin{figure}[h]
	\begin{center}
		\includegraphics[scale=0.5]{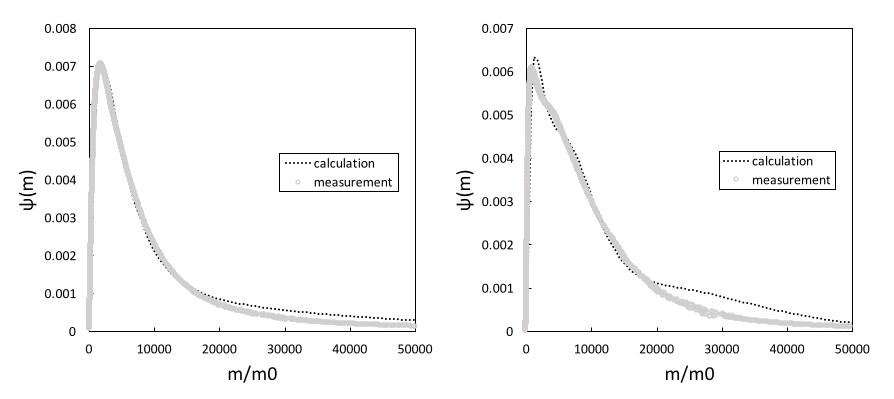}
		\caption{Multimodal MMD calculation in comparison with HT-GPC measurement for (a) grade 2 with $m_0=42.08$ Da and (b) grade 3 with $m_0=28.05$ Da.}
		\label{fig4}
	\end{center}
\end{figure}

In figure~\ref{fig4}, the MMDs of the remaining two polymers, grade 2 and 3, are plotted together with the HT-GPC measurements. Both are in good overall agreement with the light scattering evaluation, especially if the results are compared to the original Thimm et al. method \cite{Thim1999}, as is discussed in section~\ref{5.4}. The number of GE modes for calculating a bimodal MMD has to be equal to 2 or larger and was chosen to be 4 in this case. This ensures a better result, such as in the unimodal cases.

\begin{figure}[h]
	\begin{center}
		\includegraphics[scale=0.5]{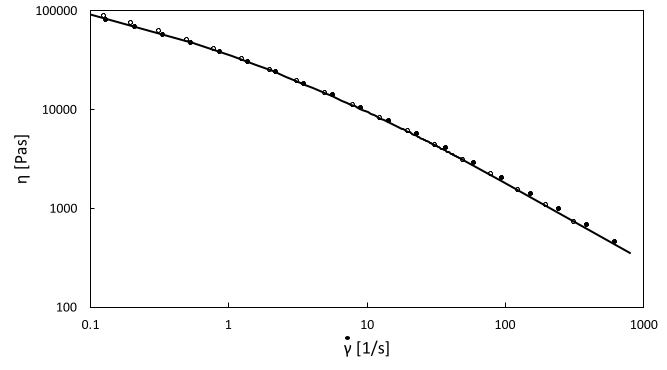}
		\caption{Viscosity curves of grade 1 from Cox-Merz rule (open circles) and measurement (full dots) compared to model fit.}
		\label{fig5}
	\end{center}
\end{figure}

\begin{figure}[h]
	\begin{center}
		\includegraphics[scale=0.5]{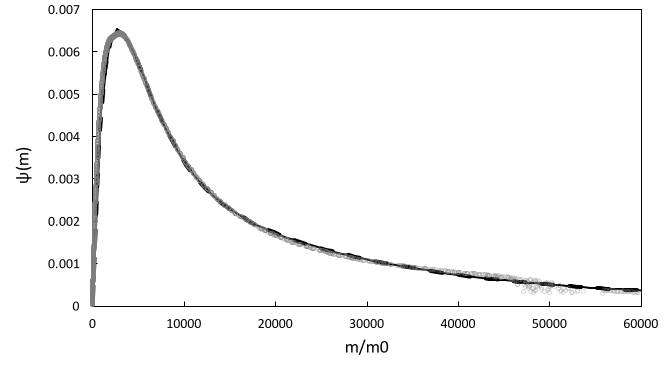}
		\caption{Trimodal MMD calculation in comparison with HT-GPC measurement for grade 1 with $m_0=42.08$ Da from viscosity curve (broad dashed line) and moduli (thin full line).}
		\label{fig7}
	\end{center}
\end{figure}

The measured nonlinear viscosity curve for grade 1 is compared to the quasi-nonlinear form of the linearized rheological measurements in figure~\ref{fig5}. The respective Cox-Merz shifting was performed with a shifting coefficient $A=1/2$ for the shear rate. The same shifting was used after approximation of the quasi-nonlinear viscosity curve by equation~\ref{viscosity}, which is also shown in figure~\ref{fig5}. It is seen that the approximation is very good. In figure~\ref{fig7}, the resulting MMDs from the moduli fit as well as the viscosity fit are compared. It can be seen that the differences between the curves are negligible.

\section{\label{5} Discussion}

\subsection{\label{5.1} Polymer Dynamics}
In the proposed model, the fractal reptation process is implicitly assumed. This means that the reptation inside the tube as well as the constraint release process are regarded as nonlinearly superimposed single Doi and Edwards \cite{DoEd1986} reptation processes. For many polymers, this form is already describing the macroscopic behavior well for a nonfractal exponent of $\beta=2$ and reproduces the macroscopic behavior to a sufficient accuracy \cite{deCl1988}, \cite{Thim2000}. However, the behavior of certain polymers can be even more complicated than that which forces to use a fractal reptation exponent. For the studies here, an exponent of $\beta=3.84$ was taken in accordance with the paper by Thimm et al. \cite{Thim1999}, despite of a later remark by the same group \cite{Thim2000}, attributing this number to disregarded Rouse modes. Using the given number is reasonable, since the present method can be used properly without considering the low molecular weight contributions as is described in section~\ref{5.3}. This, however, has consequences for the resulting MMD, as is discussed in section~\ref{5.4}. The number for $\beta$ is easily found by using a unimodal GE distribution and fitting it to the HT-GPC measurement, fixing along this way the MMD related parameters and taking the coefficient $\zeta$ and exponent $\alpha$ from the literature given in table~\ref{tab2}. The fractal exponent is then found by fitting equations~\ref{gprime} and \ref{gdprime} to the measurements.
In principle, one can also conduct the procedure given above for all unknown coefficients, but this is, of course, not a option for any direct use of the MMD calculation procedure. It should be emphasized instead that measuring the needed quantities directly is the best method for a correct MMD calculation, while numbers that cannot be measured, such as the reptation exponent, have to be taken from the literature.

\subsection{\label{5.2} GE distribution}
Here, the multimodal GE distribution was used. Due to the particular form of the kernel function taken by Thimm et al. \cite{Thim1999}, it is possible to bring every mode to the Laplace integral form. Whether one uses a unimodal distribution or a multimodal is thus irrelevant for the form of the shown equations. However, the numerical experiments showed that the multimodal form is in every way giving a better result than the unimodal form.
A drawback of the multimodal GE form is that one cannot identify any physical parameters of the distribution. All numbers are thus purely mathematical. 

The important point here is that the basic equality one has to assume when equation~\ref{ThimmTheo} is compared with the GE distribution is only correct for N modes.
If the proposed method is put to use, however, it is necessary to constrain the number of modes of the GE function in order to end up with a regularized problem. Although no numerical experiments were conducted in order to identify the transition from a regularized to an ill-posed problem here, one can refer to a very similar problem in this field, namely the calculation of the relaxation time spectrum from oscillatory rheometry, where Baumgaertel and Winter \cite{BaWi1989} were able to show that a number of eight free parameters in their least squares method was the threshold to ill-posedness.
One can, therefore, expect that the threshold for this method is found around the same number of parameters. This would mean that the problem remains regularized at least up to the 4-mode GE distribution and starts to be ill-posed at an unknown number of modes above. No numerical experiments were conducted yet to explore this boundary, but repetition of the fitting process leads to the same values of all the coefficients for the numerical experiment of grade 3, shown in figure~\ref{fig4}, which had four modes.

If equation~\ref{GEX} is written in terms of molecular mass moments, like for example $m_0$, the other parameters have to be changed accordingly, in order to provide a full basis of the outlined function. This would be a possible way of further exploration of the upper boundary of the mode numbers. In order to clarify this point, the equation is rewritten for the unimodal case as:
$$\psi(m,m_0,a,b)=\frac{1}{m_0\Gamma\left(\frac{a+1}{b}\right)}\left(\frac{m}{m_0}\right)^a\exp\left[-\left(\frac{m}{m_0}\right)^b\right]~~,$$
where it can be seen that the parameters used in equation~\ref{GEX} now have interdependencies. In particular, $\gamma=1/\Gamma(a+1/b)m_0^{a+1}$ is connected with $c=m_0^{-b}$. The change of notation allows for further insight into the resulting parameters of a fitting operation, but it is to be expected that the choice of one particular MMD moment does not provide the best fit to the actual MMD. One can now expand this form to N modes by multiplying each term of the equation with a weight $w$, s.t.:
$$\psi(m)=w_0\psi(m,m_0,a_0,b_0)+w_1\psi(m,m_1,a_1,b_1)+ . . .=\sum_{i=0}^{N}w_i\psi(m,m_i,a_i,b_i)~~,$$
with the condition that:
$$1=\sum_{i=0}^Nw_i~~.$$
By rewriting all occurring coefficients:
$$\gamma_i=\frac{w_i}{m_i^{a_i+1}\Gamma(\frac{a_i+1}{b_i})}~~,$$
the original equation~\ref{GEX} is retained. The advantage of this form would be that additional modes are identified by their weight and contribute accordingly to the overall function. Additionally, one could include an error limit $\epsilon\geq|\sum_{i=0}^{N-1}\psi(m,m_{N-1},a_{N-1},b_{N-1})-\psi(m)|$ for which a further contribution is negligible, thus identifying the upper limit of modes.

\subsection{\label{5.3}Viscosity}

Although the method of calculation for the quasi-nonlinear viscosity is based on the validity of the Cox-Merz rule, which is assumed here on the basis of a paper by Winter \cite{Wint2009}, the results fit the nonlinear measurements very well. The application of equation~\ref{viscosity} will mostly need to be complemented with the correct Cox-Merz coefficient, which shall be called the shifting parameter hereafter.

It is  indispensable to obtain the shifting parameter before applying the given method in order to obtain a mmd that is quantitatively comparable to a classical light-scattering evaluation. A qualitative mmd can be obtained without knowing the shifting parameter. 
 If the full data set for nonlinear and linearized rheology is available, it is recommended to calculate the shifting parameter from the data before applying the method. This, of course, makes the viscosity calculation of the MMD useless, since it is completely redundant. The method starting from linearized rheological data alone would completely solve the problem at hand.
 
 However, if the method starts from a nonlinear rheological measurement alone and it is used e.g. in order to trace the process related changes of the mmd of a given grade, the method can be used to detect deviations from the starting mmd. A qualitative comparison, in this case, is satisfying and the shifting parameter is not needed. 
 If the Cox-Merz coefficient of a specific grade is known from previous measurements, the method will be able to give quantitative results. 

A comparison of the given method to the method by Malkin and Theishev \cite{MaTe1987} which was further developed by Nobile et al. \cite{NoCo+1996} is interesting from a theoretical point of view. Their viscosity curve can be parametrized in GE modes as well, leading to a comparable result in terms of incomplete Laplace integrals:
\begin{equation}
\label{NoCo}
\eta(\dot{\gamma})=\eta_0\left\{1+\frac{1}{N}\sum_{i=1}^N\mathcal{L}_{\nu_i}\left(b_i,K^{c_i}\right)\left[f(\dot{\gamma},m)\right]\right\}^\alpha~~,
\end{equation}
where $\nu=\frac{a+2}{c}$, $K=\frac{1}{m_0}\left(\frac{B}{\dot{\gamma}}\right)^{1/\alpha}$, B is the relaxation time of the polymer melt as defined by Carreau\cite{Carr1972} and 
\begin{equation}
\label{NoCoFunction}
f(\dot{\gamma},m)=Kb^{\nu-1/c}m^{-1}-b^\nu
\end{equation}
The proposed model, on the other hand, is given as:
\begin{equation}
\label{MyFunction}
|\eta^\ast(\omega)|=\sum_{i=1}^N\frac{\hat{\lambda}_i}{\mu_i}\left[\Gamma^2(\hat{\xi})\mathcal{L}^2_{\hat{\xi}}(k_i,0)[\overline{f}'(\tau,\omega)+\Gamma^2(\hat{\xi})\mathcal{L}^2_{\hat{\xi}}(k_i,0)[\overline{f}''(\tau,\omega)]\right]^{1/2}~~,
\end{equation}
where $\overline{f}'=\hat{f}'/\omega^2$ and $\overline{f}''=\hat{f}''/\omega^2$.
It is seen that the largest differences between the two models are in the order of their exponent, which is considerably higher in the theory by Nobile et al. \cite{NoCo+1996}, as well as in the lower molecular weight cutoff which is zero in equation~\ref{MyFunction}.

However, it needs to be said that it seems to be a matter of taste if the low molecular weight cutoff, associated with Rouse like modes, is accounted for in the given procedure or not, since the obtained molecular mass distribution was the same for all numerical experiments conducted here whether or not the cutoff was considered. It is, nonetheless, important to notice that a change in the reptation coefficient could be necessary when the Rouse modes are neglected. This in turn would change the predictions for high molecular weight components, as is discussed in section~\ref{5.4}.  The given equations in terms of incomplete Laplace transforms can, nonetheless, fully account for the incorporation of a cutoff and it remains a possibility that the lower molecular weights have to be excluded for certain cases. Also from a theoretical point of view, the cutoff seems to be a necessity. It is, therefore, appended here that a cutoff can be implemented in equation~\ref{ShearModulus} by replacing the zero with $\tau_e^{-1}$, where $\tau_e$ corresponds to the molecular weight of entanglement $m_e$, and similarly one can implement a cutoff for equations~\ref{gprime} and \ref{gdprime} by replacing the zero with $\tau_e^\mu$.

The two theoretical approaches are not largely different in other aspects and it could be of interest to investigate the consequences of this relation more closely. 

\subsection{\label{5.4}Agreement with the measurement}

From figure~\ref{fig4} and \ref{fig6} it is seen that the overall agreement of the given method with a light scattering evaluation in HT-GPC measurements is good, especially for simple grades such as grade 1 and 2. A review of 5 models from the literature \cite{BeSl1977}, \cite{MaTe1988}, \cite{NoCo+1996}, \cite{CaGu1997}, \cite{Thim1999}, concerning efficiency as well as accuracy in MMD calculation, has shown \cite{Lang2016} that the Thimm et al. method was the most accurate and reliable. Therefore, results of the method presented here will be compared to those of the original method by Thimm et al. \cite{Thim1999}. It is seen that the mean square deviation of the calculated MMD from the HT-GPC measurement has been reduced by a factor 10 or more by  using the 4 mode Laplace method in the case of unimodal polymers, depending on the grade. A list of the mean square deviations together with their factor of improvement is given in table~\ref{tab3} for all grades. Thereby, the high molecular weight tail of grade 2 is not reproduced with the same accuracy as that of grade 1, giving a larger mean square deviation. The reason behind this could be the overweighting of high molecular weight components resulting from the value of the reptation exponent $\beta$. Similar results have been shown by Nobile et al. \cite{CoNo2003} who attribute this overweighting to a neglect of the Rouse spectrum, resulting in values of the reptation exponent significantly higher than 2. 
In the case of grade 3, the agreement of curves is only qualitative, since all important features are covered by the given method but the peak hights and the high molecular weight tail are not correctly reproduced. Hence, also the improvement regarding accuracy of the calculated mmd as shown in table~\ref{tab3} is low. 

\begin{table}[h]
	\caption{Mean square deviation of 4 mode Laplace method (msdL) and Thimm et al. method (msdT) for grades 1 to 3.}
	\smallskip{}
	\centering{}
	\begin{tabular}{||c|c|c|c||}
		\hline
		\textbf{grade} & \textbf{msdL} & \textbf{msdT} & \textbf{factor}\\
		\hline
		1 &0.000271& 0.002703 & 9.97 \\
		2 & 0.003681& 0.042883 & 11.65 \\
		3 & 0.083645 & 0.124876 & 1.49 \\
		\hline
	\end{tabular}
	\label{tab3}
\end{table}

This could be an indication of the limits of the given method. In the case of grade 3, a reason that the agreement of measurement and calculation was unsatisfactory could be that grade 3 is an ethylene 1-octene copolymer with a low degree of chain branching. The given method looses its validity for the case of chain branching, since the relaxation time dependence of the molecular mass breaks down and the relaxation process cannot be accounted for in the given manner. Although the level of chain branching is low, the individual mass associated with branches adds up to the second peak appearing in the mmd. Since this feature is apparently not fully identified from the rheometric response, the underestimation of intermediate masses adds up to the overweighting of high masses which results from the neglected Rouse modes.
One can, therefore, safely use the outlined method only for the case of mostly linear polydisperse polymers, where the given set of equations holds.

In addition, it is seen from figure~\ref{fig6} that it seems to be irrelevant which set of rheometrical measurements is taken to calculate the molecular mass distribution for the case of linear polydisperse polymers. The same was found for all grades under inspection. The use of this fact, nonetheless, clearly depends on the availability of a Cox-Merz shifting coefficient, as has been already discussed in section~\ref{5.3}.

\section{Conclusions}
\label{6}

The presented method for MMD calculation from MPs is very easy to implement and resembles in large part the well-known methods for HT-GPC detection. The method is based on the modern understanding of polymer dynamics but takes account for processes which occur on top of single reptation only in form of a non-linear superposition of simple processes, termed fractal reptation. The use of a multimodal GE function allows for the given form of Laplace integrals but does not give rise to a straight forward interpretation of the distribution in terms of physical quantities. 
It is shown that the given approach yields good agreement with HT-GPC measurements for the case of linear polydisperse polymers. Occurring material constants, however, have to be measured a priori in oder to use the method to its full potential. In the case of high chain branching or immiscibility of blends, the method seems to loose its validity. Additionally, discrepancies with the measurement in the high molecular mass regime are encountered. These can be attributed to the neglect of Rouse modes. Nonetheless, bimodal or higher modal distributions can be calculated with equal results from either MP chosen, as long as the Cox-Merz rule remains valid.

\section*{Acknowledgments}

The author wishes to acknowledge M. Aigner from Erema GmbH , T. K\"opplmayr from Engel GmbH and J. Miethlinger from the Institute of Polymer Extrusion and Building Physics at JKU Linz for their help in conducting rheological measurements and the fruitful discussions and I. Teasdale from the Institute of Polymer Chemistry at JKU Linz for his help in conducting and evaluating HT-GPC measurements. This research is funded by the European Union within the Horizon 2020 project under the DiStruc Marie Sk\l{}odowska Curie innovative training network; Grant Agreement No. 641839.


%
%

%


\bibliographystyle{unsrtnat}
\nocite{*}
\bibliography{LapTransAr}

\providecommand{\noopsort}[1]{}\providecommand{\singleletter}[1]{#1}%
\begin{thebibliography}{42}
\providecommand{\natexlab}[1]{#1}
\providecommand{\url}[1]{\texttt{#1}}
\expandafter\ifx\csname urlstyle\endcsname\relax
  \providecommand{\doi}[1]{doi: #1}\else
  \providecommand{\doi}{doi: \begingroup \urlstyle{rm}\Url}\fi

\bibitem[Evans and Morriss(2008)]{EvMo2008}
D.~J. Evans and G.~Morriss.
\newblock \emph{Statistical Mechanics of Nonequilibrium Liquids}.
\newblock Cambridge, 2008.

\bibitem[Doi and Edwards(1986)]{DoEd1986}
M.~Doi and S.~F. Edwards.
\newblock \emph{The theory of polymer dynamics}.
\newblock Oxford, 1986.

\bibitem[Anderssen and Mead(1998)]{Ande1998}
R.~S. Anderssen and D.~W. Mead.
\newblock Theoretical derivation of molecular weight scaling for rheological
  parameters.
\newblock \emph{J. \ Non-Newtonian Fluid \ Mech.}, 76:\penalty0 299, 1998.

\bibitem[Thimm et~al.(1999)Thimm, Friedrich, Marth, and Honerkamp]{Thim1999}
W.~Thimm, C.~Friedrich, M.~Marth, and J.~Honerkamp.
\newblock An analytical relation between relaxation time spectrum and molecular
  weight distribution.
\newblock \emph{J. \ Rheol.}, 43:\penalty0 1663, 1999.

\bibitem[Carrot and Guillet(1997)]{CaGu1997}
C.~Carrot and J.~Guillet.
\newblock From dynamic moduli to molecular weight distribution: A study of
  various polydisperse linear polymers.
\newblock \emph{J. \ Rheol.}, 41:\penalty0 1203, 1997.

\bibitem[Nobile and Cocchini(2001)]{NoCo2001}
M.~R. Nobile and F.~Cocchini.
\newblock Evaluation of molecular weight distribution from dynamic moduli.
\newblock \emph{J. \ Rheol.}, 40:\penalty0 111, 2001.

\bibitem[Nobile et~al.(1996)Nobile, Cocchini, and Lawler]{NoCo+1996}
M.~R. Nobile, F.~Cocchini, and J.~V. Lawler.
\newblock On the stability of molecular weight distributions as computed from
  the flow curves of polymer melts.
\newblock \emph{J. \ Rheol.}, 40:\penalty0 363, 1996.

\bibitem[Malkin and Theishev(1987)]{MaTe1987}
A.~Y. Malkin and A.Y. Theishev.
\newblock Can the mmd of a polymer be determined uniquely from the flowcurve of
  its melt?
\newblock \emph{Polym. \ Sci. \ USSR}, 29:\penalty0 2449, 1987.

\bibitem[Malkin and Theishev(1988)]{MaTe1988}
A.~Y. Malkin and A.Y. Theishev.
\newblock Method of determining molar mass distribution from curves of polymer
  melt flow.
\newblock \emph{Polym. \ Sci. \ USSR}, 30:\penalty0 195, 1988.

\bibitem[van Ruymbeke et~al.(2005)van Ruymbeke, Keunings, and
  Bailly]{RuKe+2005}
E.~van Ruymbeke, R.~Keunings, and C.~Bailly.
\newblock Prediction of linear viscoelastic properties for polydisperse
  mixtures of entangled star and linear polymers: Modified tube-based model and
  comparison with exerimental results.
\newblock \emph{J. \ Non-Newtonian\ Fluid \ Mech.}, 128:\penalty0 7, 2005.

\bibitem[Pattamaprom et~al.(2008)Pattamaprom, Larson, and Sirivat]{PaLa+2008}
C.~Pattamaprom, R.~G. Larson, and A.~Sirivat.
\newblock Determining polymer molecular weight distributions from rheological
  properties using the dual-constraint model.
\newblock \emph{Rheol. \ Acta}, 47:\penalty0 689, 2008.

\bibitem[Greschner(1981)]{Gresc1981}
G.~S. Greschner.
\newblock \emph{Maxwellgleichungen - Bd. 2}.
\newblock Huethig und Wepf, 1981.

\bibitem[Deiber et~al.(1993)Deiber, Peirotti, and Bortolozzi]{DePe+1993}
J.~A. Deiber, M.~B. Peirotti, and J.~R. Bortolozzi.
\newblock Estimation of molecular weight distributions of elastomers and
  polymer melts through dynamic rheometry.
\newblock \emph{J. \ Elastomers and Plastics}, 25:\penalty0 22, 1993.

\bibitem[Ressia et~al.(2000)Ressia, Villar, and Vales]{ReVi+2000}
J.~A. Ressia, M.~A. Villar, and E.~M. Vales.
\newblock Influence of polydispersity on the viscoelastic properties of linear
  polydimethylsiloxanes and their binary blends.
\newblock \emph{Polymer}, 41:\penalty0 6885–6894, 2000.

\bibitem[Grossiord et~al.(1988)Grossiord, Couarraze, and Leclerc.]{Gros1988}
J.~L. Grossiord, G.~Couarraze, and B.~Leclerc.
\newblock New methods of determination of the molecular weight distribution
  from rheological measurements.
\newblock \emph{Rheol. \ Acta}, 27:\penalty0 487, 1988.

\bibitem[Cocchini and Nobile(2003)]{CoNo2003}
F.~Cocchini and M.~R. Nobile.
\newblock Constrained inversion of rheological data to molecular weight
  distribution for polymer melts.
\newblock \emph{Rheol. \ Acta}, 42:\penalty0 232, 2003.

\bibitem[Vaidya and Hester(1984)]{VaHe1984}
R.~A. Vaidya and R.~D. Hester.
\newblock Deconvolution of overlapping chromatographic peaks using constrained
  non-linear optimization.
\newblock \emph{J. \ Chromat.}, 287:\penalty0 231, 1984.

\bibitem[Temme(1987)]{Temm1987}
N.~M. Temme.
\newblock Incomplete laplace integrals: uniform asymptotic expansion with
  application to the incomplete beta function.
\newblock \emph{SIAM \ J. \ Math. \ Anal.}, 18:\penalty0 1638, 1987.

\bibitem[Bor(2016)]{Bore0000}
www.borealisgroup.com/en/polyolefins/products, 2016.

\bibitem[Dow(2016)]{Dow0000}
http://www.dow.com/plasticpipes/eu/products/dowlex/2388pe.htm, 2016.

\bibitem[Beer et~al.(2001)Beer, Capaccio, and Rose]{BeCa+2000}
F.~Beer, G.~Capaccio, and L.~J. Rose.
\newblock High molecular weight tail and long-chain branching in low-density
  polyethyenes.
\newblock \emph{J. \ Appl. \ Polym. \ Sci.}, 80:\penalty0 2815, 2001.

\bibitem[The(2016)]{Ther2016}
www.thermofisher.com/order/catalog/product/567-2020, 2016.

\bibitem[Cole and Cole(1942)]{CoCo1942}
K.~S. Cole and R.~H. Cole.
\newblock Dispersion and absorption in dielectrics ii. direct current
  characteristics.
\newblock \emph{J. \ Chem. \ Phys.}, 10:\penalty0 98, 1942.

\bibitem[Lang(2015)]{Lang2016}
C.~Lang.
\newblock \emph{MMD of polydisperse linear polymers from rheological data}.
\newblock Akademiker Verlag, 2015.

\bibitem[Schausbeger(1991)]{Scha1991}
A.~Schausbeger.
\newblock A description of the linear viscoelasticity of molten linear
  monodisperse polystyrenes with the aid of a generalized discrete relaxation
  time spectrum.
\newblock \emph{Rheol \ Acta}, 30:\penalty0 197, 1991.

\bibitem[Eder et~al.(1989{\natexlab{a}})Eder, Janeschitz-Kriegl, Liedauer,
  Schausberger, Stadlbauer, and Schindlauer]{Eder1989}
G.~Eder, H.~Janeschitz-Kriegl, S.~Liedauer, A.~Schausberger, W.~Stadlbauer, and
  G.~Schindlauer.
\newblock The influence of molar mass distribution on the complex moduli of
  polymer melts.
\newblock \emph{J. \ Rheol.}, 33:\penalty0 805, 1989{\natexlab{a}}.

\bibitem[Levenberg(1944)]{Leve1944}
K.~Levenberg.
\newblock A method for the solution of certain non-linear problems in least
  squares.
\newblock \emph{Quart. \ Appl. \ Math.}, 2:\penalty0 164, 1944.

\bibitem[Marquardt(1963)]{Marq1963}
D.~Marquardt.
\newblock An algorithm for least-squares estimation of nonlinear parameters.
\newblock \emph{SIAM \ J. \ Appl. \ Math.}, 11:\penalty0 431, 1963.

\bibitem[des Cloizeaux(1988)]{deCl1988}
J.~des Cloizeaux.
\newblock Double repattion vs. simple reptation in polymer melts.
\newblock \emph{Europhys. \ Lett.}, 5:\penalty0 437, 1988.

\bibitem[Thimm et~al.(2000)Thimm, Friedrich, and Marth]{Thim2000}
W.~Thimm, C.~Friedrich, and M.~Marth.
\newblock On the rouse spectrum and the determination of the molecular weight
  distribution from rheological data.
\newblock \emph{J. \ Rheol.}, 44:\penalty0 429, 2000.

\bibitem[Baumgaertel and Winter(1989)]{BaWi1989}
M.~Baumgaertel and H.~H. Winter.
\newblock Determination of discrete relaxation and retardation time spectra
  from dynamic mechanical data.
\newblock \emph{Rheol. \ Acta}, 28:\penalty0 519, 1989.

\bibitem[Winter(2009)]{Wint2009}
H.~H. Winter.
\newblock Three views of viscoelasticity for cox-merz materials.
\newblock \emph{Rheol. \ Acta}, 48:\penalty0 241, 2009.

\bibitem[Carreau(1972)]{Carr1972}
P.~J. Carreau.
\newblock Rheological equations from molecular network theories.
\newblock \emph{Trans. \ Soc. \ Rheol.}, 16:\penalty0 99, 1972.

\bibitem[Bersted and Slee(1977)]{BeSl1977}
B.~H. Bersted and J.~D. Slee.
\newblock A relationship between steady-state shear melt viscosity and
  molecular weight distribution in polystyrene.
\newblock \emph{J. \ Appl. \ Polym. \ Sci.}, 21:\penalty0 2631, 1977.

\bibitem[H.Watanabe et~al.(2004)H.Watanabe, Ishida, Matsumyia, and
  Inoue]{WaIs+2004}
H.Watanabe, S.~Ishida, Y.~Matsumyia, and T.~Inoue.
\newblock Test of full and partial tube dilation pictures in entangled blends
  of linear polyisoprenes.
\newblock \emph{Macrom.}, 37:\penalty0 1937, 2004.

\bibitem[Ferry(1980)]{Ferr1980}
J.~D. Ferry.
\newblock \emph{Viscoelastic properties of polymers}.
\newblock Wiley, 1980.

\bibitem[Moiseiwitsch(1977)]{Mois1977}
B.~L. Moiseiwitsch.
\newblock \emph{Integral Equations}.
\newblock Longman, 1977.

\bibitem[Dealy and Wissbrun(1990)]{DeWi1990}
J.~M. Dealy and K.~F. Wissbrun.
\newblock \emph{Melt Rheology and its role in plastics processing}.
\newblock Van Nostrand, 1990.

\bibitem[Mead(1994)]{Mead1994}
D.~W. Mead.
\newblock Determination of molecular weight distributions of linear flexible
  polymers from linear viscoelastic material functions.
\newblock \emph{J. \ Rheol.}, 38:\penalty0 1797, 1994.

\bibitem[van Ruymbeke et~al.(2010)van Ruymbeke, Coppola, Balacca, Righi, and
  Vlassopoulos]{vaRu2010}
E.~van Ruymbeke, S.~Coppola, L.~Balacca, S.~Righi, and D.~Vlassopoulos.
\newblock Decoding the viscoelastic response of polydisperse star/linear
  polymer blends.
\newblock \emph{J. \ Rheol.}, 54:\penalty0 507, 2010.

\bibitem[van Ruymbeke et~al.(2009)van Ruymbeke, Vlassopoulos, Kapnistos, Liu,
  and Bailly]{RuVl+2010}
E.~van Ruymbeke, D.~Vlassopoulos, M.~Kapnistos, C.~Y. Liu, and C.~Bailly.
\newblock Proposal to resolve the time-stress discrepancy of tube models.
\newblock \emph{Macrom.}, 43:\penalty0 525, 2009.

\bibitem[Eder et~al.(1989{\natexlab{b}})Eder, Janeschitz-Kriegel, Liedauer,
  Schausberger, Stadlauer, and Schindlauer]{EdJa+1989}
G.~Eder, H.~Janeschitz-Kriegel, S.~Liedauer, A.~Schausberger, W.~Stadlauer, and
  G.~Schindlauer.
\newblock The influence of molar mass distribution on the complex moduli of
  polymer melts.
\newblock \emph{J. \ Rheol.}, 33:\penalty0 805, 1989{\natexlab{b}}.

\end{thebibliography}

\end{document}